	\documentclass[journal,comsoc]{IEEEtran}
\IEEEoverridecommandlockouts
\usepackage{amsmath,amssymb,amsfonts}
\usepackage{algorithmic}
\usepackage{graphicx}
\usepackage{textcomp}
\usepackage{authblk}
\usepackage{enumitem}
\usepackage{multicol}

\graphicspath{ {Figures/} }
\def\BibTeX{{\rm B\kern-.05em{\sc i\kern-.025em b}\kern-.08em
		T\kern-.1667em\lower.7ex\hbox{E}\kern-.125emX}}

\begin{document}
		\title{Design of Low-Complexity Convolutional Codes over GF$(q)$}
		\author[1]{Rami~Klaimi}
		\author[1]{Charbel~Abdel Nour}
		\author[1]{Catherine~Douillard}	
		\author[2]{Joumana~Farah}
		
		\affil[1]{ IMT Atlantique,  Lab-STICC, UBL, F-29238 Brest, France }
		\affil[2]{ Department of Electricity and Electronics, Faculty of Engineering\protect\\ Lebanese University, Roumieh, Lebanon }

	\markboth{Accepted for publication in IEEE-Globecom 2018}%
	{}
	
		\maketitle
	
	\begin{abstract}

This paper proposes a new family of recursive systematic convolutional codes, defined in the non-binary domain over different Galois fields GF$(q)$ and intended to be used as component codes for the design of non-binary turbo codes. A general framework for the design of the best codes over different GF$(q)$ is described. 
The designed codes offer better performance than the non-binary convolutional codes found in the literature. They also outperform their binary counterparts when combined with their corresponding QAM modulation or with lower order modulations.

	\end{abstract}
	\begin{IEEEkeywords}
Finite field, non-binary codes, recursive systematic convolutional codes,  coded modulation, turbo codes.
	\end{IEEEkeywords}
	
	\section{Introduction}
	
Since the early 2000s, binary turbo and low density parity-check (LDPC) codes \cite{TurboCodeBerrou,LDPCGallager,LDPCMacKay}, have been adopted in many communication standards, such as the third, fourth and fifth generations of mobile communications (3G, 4G, 5G), the second generation of digital video broadcasting (DVB) and the WiMAX standards. When transmitting long codewords, most of these codes are known to approach the Gaussian channel capacity very closely. However, they do not perform so close to the theoretical limit when small block lengths are considered \cite{liva2016code}. The performance loss is due to the correlation experienced in the iterative decoding process of short data blocks \cite{CorrelationHokfelt}.

These observations have triggered much attention from the channel coding research community, and are being taken into consideration with the increasing need for short packet transmission, for instance  for machine to machine communications. In particular, numerous studies have investigated the design of codes over high-order Galois fields (GF), especially for the LDPC family, and have shown the  potential of these codes \cite{MacKayNBLDPC,PoulliatNBLDPC}. New structures of non-binary (NB) LDPC codes are proposed  where the encoded NB symbols are directly mapped to a NB modulation with the same order. These codes are jointly designed with the corresponding modulation depending on the GF order.

While NB LDPC codes have been widely studied in the literature, research related to NB  convolutional and turbo codes over GF$(q)$, $q>2$ is very limited. Convolutional codes over rings are defined in \cite{konishi2014coded} using a matched mapping in order to easily find the best codes. In \cite{LivaNBTC1,LivaNBTC2}, NB turbo codes are constructed derived from protograph sub-ensembles of regular LDPC codes. These codes are defined as a concatenation of two NB time-variant accumulators. Also, convolutional codes over GF$(q)$ are defined in \cite{zhao2016convolutional}, where the author has limited the study to codes over GF$(4)$. Using the results in \cite{zhao2016convolutional}, turbo codes over GF$(4)$ are defined in \cite{zhao2016turbo}. In addition, turbo codes over GF$(4)$ with different types of channels are studied in \cite{abd2016TC1,abd2017TC2}. A general study for the design of convolutional codes to be used as component codes for NB turbo codes over different GF$(q)$ seems to be missing in the literature.

\begin{figure}[htbp]
		\centering
		\includegraphics[width=0.48\textwidth]{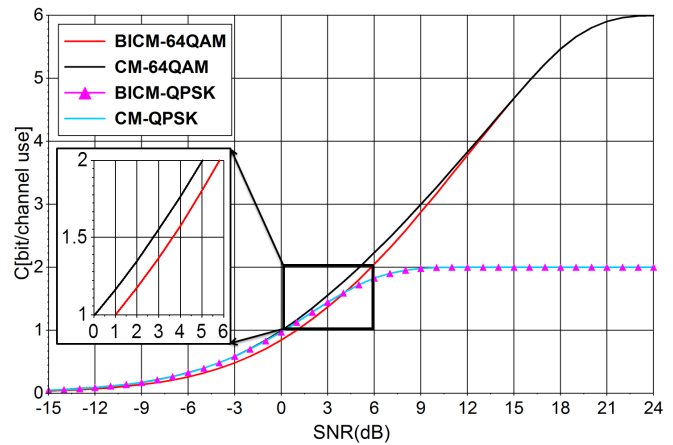}
		\caption{BICM and CM capacities versus signal-to-noise ratio (SNR) over an AWGN channel.}
		\label{CapacityFig}
	\end{figure}

From an information theoretic perspective, previous studies have shown that bit-interleaved coded modulation (BICM) schemes suffer from a capacity loss compared to coded modulation (CM) schemes \cite{AlvaradoCMBICM,CaireCMBICM}. This loss is even more pronounced for high modulation orders and at low spectral efficiencies. An example of comparison between the CM and BICM capacities is illustrated in Fig.~\ref{CapacityFig}, where a transmission with 64-QAM and QPSK modulations over an additive white Gaussian noise (AWGN) channel is considered. The BICM and CM capacities are calculated based on the model presented in \cite{CaireCMBICM}.
	
Fig.~\ref{CapacityFig} shows a gain in capacity of more than 1.0 dB for the coded modulation using 64-QAM for spectral efficiencies lower than 2 bit/channel use. In other words, a potential gain of more than 1.0 dB of SNR can be achieved by using a NB coding scheme with coding rate 1/3  over GF(64) combined with a 64-QAM modulation, compared to a BICM scheme using binary coding. This is mainly due to the fact that, in the NB structure, the encoded symbols are directly mapped onto the modulation, instead of being marginalized as in the binary case. However, Fig.~\ref{CapacityFig} shows almost equal CM and BICM capacities in the case of QPSK for all SNR regions. Therefore, from the capacity standpoint, there is no real interest in using codes over GF(4) combined with QPSK modulation.
	
From these results, one can assume that well-designed NB codes can outperform their binary counterparts for certain coding rate ranges and modulation orders. 
To this end, this paper presents a detailed study of the design of convolutional codes over NB GF$(q)$, to be used as constituent codes for NB turbo codes.

This contribution is structured as follows: in Section \ref{sec:CodeStructure} we propose a new recursive systematic convolutional code structure over GF$(q)$ and we show its superiority with respect to the classical structure with the same number of states. Section \ref{sec:Choice} starts by describing the used design criterion, then evaluates the impact of the constellation mapping on the search procedure. The used decoding algorithm, which is the extension of the Max-Log-MAP algorithm to the symbol domain, is explained in Section \ref{sec:MinLogMAPNB}. Simulation results comparing the proposed codes with the best published binary and non-binary convolutional codes in the state of the art are provided in Section \ref{sec:Results}. Section \ref{sec:Disc} concludes the paper.
	
	\section{Non-binary recursive systematic convolutional code structures}\label{sec:CodeStructure}
\begin{figure}[ht]
		\centering
		\includegraphics[width=0.4\textwidth]{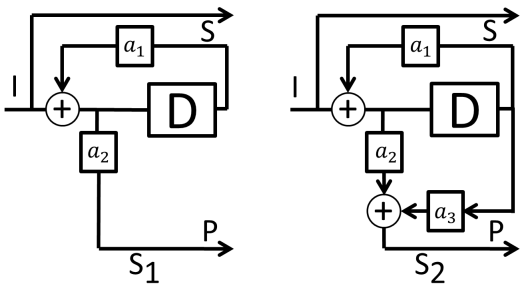}
		\caption{Non-binary convolutional code structures with one memory element.}
		\label{Fig:GFConvCode}
	\end{figure}

In order to limit the complexity of the design process, only rate-$1/2$ convolutional codes with one memory element were considered in this work. First, the accumulator structure (structure $S_{1}$ in Fig. \ref{Fig:GFConvCode}), inspired from the binary case, was used as an encoding template. Then, the proposed structure $S_{2}$ depicted in Fig. \ref{Fig:GFConvCode} was adopted for the reasons that will be presented shortly. 

	Both structures consist of a recursive systematic convolutional (RSC) code with recursion polynomial:
	\begin{equation}
	P_r(D)=1+a_1 D
	\label{Eqn:rec}
	\end{equation}
	Note that the recursive nature of the code requires $a_1 \neq 0$. 
	The structures in Fig. \ref{Fig:GFConvCode} differ in the parity polynomial. For structure $S_{1}$, it is equal to: 
	\begin{equation}
	P_{p_1}(D)=a_2+a_2a_1 D 
	\end{equation}
	and for structure $S_{2}$ to:
	\begin{equation}
	\begin{gathered}
	P_{p_2}(D) =a_2 (1+a_1 D)+ a_3 D \\
	=a_2+\left(a_1a_2+a_3\right )D  	
	\label{Eq:Parity2}
	\end{gathered}
	\end{equation}
In order for structures $S_{1}$ and $S_{2}$ to represent a convolutional code, an additional constraint should be met in every structure. For structure $S_{1}$, it is sufficient to assume that $a_2 \neq 0$. For $S_{2}$, we should have that:
	\begin{equation}
	a_1a_2+ a_3 \neq 0
	\label{eqn:cond2}
	\end{equation}
	When $a_1a_2+ a_3 = 0$, the input data is just multiplied by a fixed coefficient without being encoded. 
	
For a code defined over GF$(q)$, coefficients $a_j $, $j=\{1,2,3\}$, the input data,  systematic and parity output data are elements of GF$(q)$.	
Since the encoding structures use only one memory element, the codes considered here have the lowest level of complexity, for a given value of $q$, among all the codes over GF$(q)$. Their trellis diagrams have a number of states equal to  the order of the Galois field and are fully connected.

	Structure $S_{2}$ is more general than  structure $S_{1}$ since the latter can be derived from the former by using the same coefficient $a_1 $ and $a_2 $ and by setting $a_3=0 $.

Let $E_i$ and $E_{i+1}$ be the encoder states at time $i$ and $i+1$ and $s_i$ and $p_i$ be the systematic and parity symbols labeling the transition between $E_i$ and $E_{i+1}$. The relation between the successive encoder states is, for both structures:
	
	\begin{equation}
	E_{i+1}=s_i + a_1 E_i
	\label{EqnRelStates}
	\end{equation}
	
For structure $S_1$, the parity symbol labeling the transition between $E_i$ and $E_{i+1}$ can be written as:
	
	\begin{equation}
	p_{1,i}=a_2 \left(s_i + a_1 E_i \right) = a_2 E_{i+1}
	\label{EqnParityP1}
	\end{equation}
	
For structure $S_2$, it is written as:
	
	\begin{equation}
	p_{2,i}=a_2 \left(s_i + a_1 E_i \right) + a_3 E_i= a_2 E_{i+1} + a_3 E_i
	\label{EqnParityP2}
	\end{equation}
	
If we assume that coefficients $a_1$, $a_2$ and $a_3$ are non-zero, $(q-1)^2$ different NB convolutional codes can be obtained with structure $S_1$ when varying the coefficients, while structure $S_2$ can provide $(q-1)^3$ different codes.
For both structures, (\ref{EqnRelStates}) shows that for a given value of $a_1$, $q$ transitions stem from each state $E_i$, ending  in $q$ different possible states $E_{i+1}$. Equation (\ref{EqnRelStates}) also shows that these $q$ transitions are labeled with $q$ different values of systematic symbols. Equations (\ref{EqnParityP1}) or (\ref{EqnParityP2}) provide the corresponding $q$ different values of parity symbols for the two structures considered in this paper.
The main difference between structures $S_1$ and $S_2$ is that the parity symbol  $p_{1,i}$ is the same for the $q$ transitions arriving at state $E_{i+1}$ in $S_1$ due to (\ref{EqnParityP1}). On the contrary, (\ref{EqnParityP2}) shows that a non-zero value of $a_3$ allows all the transitions arriving at state $E_{i+1}$ in $S_2$ to be labeled with $q$ different parity values $p_{2,i}$. Consequently, structure $S_2$ allows the search space for the code to span all the combinations of systematic and parity symbols for each transition between two states in the trellis.

	\section{Searching for good NB convolutional codes}\label{sec:Choice}
	\subsection{Design Criterion} \label{sec:DesignCrit}
	
Going back to Ungerboeck's results on trellis codes \cite{UngerboeckDistancel}, the selection criterion for a code, when associated with a high-order modulation, is based on the Euclidean distance spectrum of the coded modulation instead of the Hamming distance spectrum of the code, when no bit interleaving is considered. 
As far as NB convolutional codes are concerned, their selection is usually based on the minimization of the average symbol error probability $P_s$, which is upper bounded by \cite{benedetto1999principles}:
	
	\begin{equation}
	P_s\leq\sum_{I\in{\mathcal{S}}} \sum_{\hat{I}\in{\mathcal{S}}} n(I,\hat{I})p(I)P(I\rightarrow\hat{I})
	\end{equation}
where $\mathcal{S}$ denotes the set of different possible sequences in the trellis, $I$ and $\hat{I}$ are the information sequences corresponding respectively to the correct path and to an erroneous path in the trellis, $p(I)$ is the probability that the source transmits sequence $I$,  $P(I\rightarrow\hat{I})$ is the pairwise error probability (PEP), that is, the probability that the decoder chooses erroneous sequence $\hat{I}$ instead of the correct transmitted sequence $I$, and $n(I,\hat{I})$ is the number of erroneous symbols due to such an error event. 
For the case of an AWGN channel, deriving the Chernoff bound for the PEP \cite{simon1985spread,divsalar1987trellis} gives:
	\begin{equation}
	P(I\rightarrow\hat{I}) \leq e^{-\frac{E_s}{4N_0}\sum_{n=1}^N \left|I_n-\hat{I}_n \right|^2}
	\end{equation}
where $E_s$	is the average energy per transmitted symbol for the considered constellation $\mathcal{C}$.
Therefore, minimizing the average symbol error probability amounts to maximizing the Euclidean distance between sequences in the trellis diagram, which is our primary objective in this study.

Contrary to binary convolutional codes, when associated with a high order modulation, the first terms of the distance spectrum of a NB code cannot be obtained by assuming that the all-zero sequence has been transmitted. This is due to the fact that common NB constellations used in most communication systems, such as high-order QAM constellations, do not have the uniform error property \cite{benedetto1999principles}. Therefore, to determine the distance spectrum of a NB convolutional code associated with a high order modulation, we have to consider all possible pairs of competing sequences, with paths diverging from a given state  in the trellis diagram and then converging again to a given state. Such sequence pairs are called DC (diverging and converging) sequences or paths within this paper. The corresponding cumulated Euclidean distance is computed as the sum of the Euclidean distances between symbols transmitted along the two DC paths. 
	
For DC paths stretching over $L$ trellis sections, the squared cumulated Euclidean distance between two DC sequences $X^1$ and $X^2$  is calculated as follows:
	\begin{equation}
	\begin{gathered}
	D_{Euc}^2= \sum_{l=1}^{L} \left(d^2(X^1_{ls}, X^2_{ls})+d^2(X^1_{lp}, X^2_{lp})\right)\\
	= \sum_{l=1}^{L} \left[(I_{X^1_{ls}}-I_{X^2_{ls}})^{2}+(Q_{X^1_{ls}}-Q_{X^2_{ls}})^{2}\right.\\
	\left.+(I_{X^1_{lp}}-I_{X^2_{lp}})^{2}+(Q_{X^1_{lp}}-Q_{X^2_{lp}})^{2}\right]
	\end{gathered}
	\label{Eqn:DEuc} 
	\end{equation}
where $X^b_{ls}$ and $X^b_{lp}$ are the systematic and parity values respectively, at trellis section $l$ in sequence $X^b$, $b=1,2$, and $I_x$ and $Q_x$ represent the in-phase and quadrature components of  constellation signal $x$. 
	
The search for good NB convolutional codes consists in searching for the set of coefficients $a_1$, $a_2$ and $a_3$ that maximize the lowest $D_{Euc}$ values while minimizing their multiplicities (i.e. the number of sequence pairs with a given distance).

	\subsection{Distance spectrum computation methodology} \label{sec:SearchMethod}
		
\begin{figure}[htbp]
		\centering
		\includegraphics[width=0.47\textwidth]{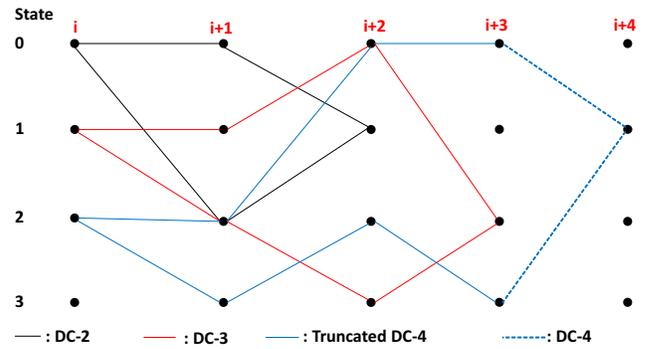}
		\caption{Examples of length-2, length-3, truncated length-4 and length-4 DC pairs of sequences in GF(4).}
		\label{DCsequences}
	\end{figure}

The distance spectrum of a NB code can be calculated by enumerating the DC sequences in the code trellis and by computing the corresponding cumulated Euclidean distances according to (\ref{Eqn:DEuc}).
 For the NB code model adopted in our study (structure $S_2$ in Fig. \ref{Fig:GFConvCode}), the trellis is fully connected. Therefore, any pair of paths in the trellis diverging from a state at time $i$ can converge to any state at time $i+2$, i.e., the shortest DC sequences have length 2.
We have observed that, whatever the values of coefficients  $a_1$, $a_2$ and $a_3$, enumerating the length-2 and length-3 DC pairs of paths (see Fig. \ref{DCsequences}) is enough to find all the sequences corresponding to the minimum cumulated Euclidean distance $d_1$ and to the second minimum cumulated Euclidean distance $d_2$ of the code. This is guaranteed since, when considering length-3 sequences diverging from a state but not converging to another (see truncated DC-4 sequences in Fig. \ref{DCsequences}), the obtained cumulated distances are greater than $d_1$ and $d_2$. 
	
Therefore,  for each examined code, we have enumerated length-2 and 3  DC sequences, computed the corresponding cumulated Euclidean distances and  stored the multiplicities for the two first distance terms  $d_1$ and $d_2$.

	\subsection{Search for convolutional codes over GF$(q)$, $q>2$, with conventional $q$-QAM constellations} \label{sec:search1}
	
As described in section \ref{sec:CodeStructure}, the code is defined through the choice of the coefficients $a_1$, $a_2$ and $a_3$. The selection of the coefficient values defining the best code depends on the mapping of the encoded symbols $X$ in GF$(q)$ to the $q$-ary constellation $\mathcal{C}$. A question that arises is whether changing the mapping has an impact on the distance spectrum of the best convolutional code that can be found by varying the coefficients $a_1$, $a_2$ and $a_3$ values. 
For a $q$-ary constellation $\mathcal{C}$, $q!$ different mappings $\mu$  can be defined. Since the values of the first and second minimum cumulated Euclidean distances, and their multiplicities, are calculated from length-2 and 3 DC sequences as stated in section \ref{sec:SearchMethod}, the study of the effect of changing the mapping $\mu$ on these sequences should be performed to answer this question. 
Two cases are considered for this study: 
\begin{itemize}
	\item The code is defined and kept unchanged (constant values are taken for the $a_j$ parameters, $j=1 \cdots 3$  ), while the constellation mapping $\mu$ spans the different $q!$ possibilities,
	\item The mapping $\mu$ is kept unchanged, while each $a_j$ code parameter spans the $q-1$ possible values.   
\end{itemize}
If both cases provide two coded modulations with the same distance spectrum,  we  can then conclude that the mapping has no impact on the result of the search procedure.
\begin{figure}[htbp] 
	\centering
	\includegraphics[width=0.3\textwidth]{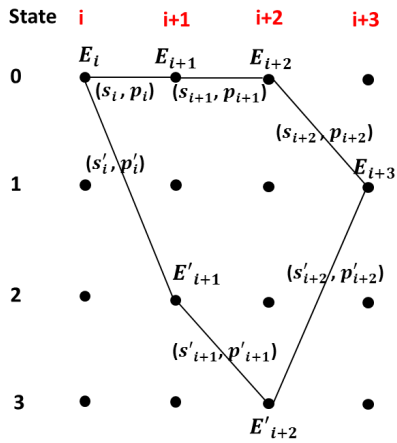}
	\caption{An example of length-3 DC sequences in a GF(4) code trellis.}
	\label{Fig:DC3Stages}
\end{figure}

Fig. \ref{Fig:DC3Stages} shows an example of two  DC sequences of length-3. The exploration of such  DC sequences can be divided into three steps: 
\begin{itemize}
	\item The first step focuses on the divergence section of the sequences, i.e. the section where they first diverge from the same state $E_i$ to two different states $E_{i+1}$ and $E'_{i+1}$. 
	\item The second step addresses the trellis section (or sections for longer sequences) where these sequences do not have any state in common. This step corresponds to the transitions from $E_{i+1}$ to $E_{i+2}$ and from $E'_{i+1}$ to $E'_{i+2}$ in the example of Fig. \ref{Fig:DC3Stages}. Note that for length-2 DC sequences, this step does not exist. 
	\item The final step concerns the trellis section where the sequences converge to the same state $E_{i+3}$. 
\end{itemize}
The exploration consists of counting  the total number of different possible cumulated Euclidean distances between DC sequences for each of the two previously mentioned study cases. It is performed for the general assumption of an arbitrarily shaped $q$-ary constellation, which represents the worst case scenario. For  QAM constellations, symmetry could be exploited to reduce the number of different possible Euclidean distances.

\paragraph{Case 1 -- $a_j$ parameters are constant and $\mu$ changes} Let $N_{\mu}$ denote the number of different possible values of the cumulated Euclidean distance between any two DC sequences of length up to 3, when $\mu$ changes. 
\begin{itemize}
	\item Step 1: The Euclidean distances are computed between transitions labeled by systematic and parity symbols emanating from the same state. They are divided into two sub-terms, one for the systematic part and one for the parity part (see (\ref{Eqn:DEuc})). The number of differently valued sub-terms for each symbol type is the number of combinations of 2 in a set of $q$ elements, denoted by $\binom{q}{2}$. When varying the mapping, any combination of any two possible values of each symbol type is possible. Therefore, the contribution of step 1 to $N_{\mu}$ is ${\binom{q}{2}}^2$. 
	\item Step 2: The Euclidean distances are computed between transitions without any state in common. Contrary to step 1, the transitions with equal values of systematic ($s_{i+1}=s'_{i+1}$) or parity ($p_{i+1}=p'_{i+1}$) symbols should also be considered. Therefore, the contribution of step 2 to $N_{\mu}$ is $\left ( {\binom{q}{2}}+1\right )^2-1$. The omitted term corresponds to the non-existing case where ($s_{i+1}=s'_{i+1}$) and ($p_{i+1}=p'_{i+1}$), due to the code structure, following (\ref{EqnRelStates}) and (\ref{EqnParityP2}).
	\item Step 3: Dictated by the code structure, by exploiting the existing symmetry with regards to step 1, we can deduce that the contribution of step 3 to $N_{\mu}$ is also ${\binom{q}{2}}^2$.   
\end{itemize}	
In summary, $N_{\mu}=\left ( \left ({\binom{q}{2}} +1\right )^2-1\right )\cdot{\binom{q}{2}}^4$.

\paragraph{Case 2 -- $\mu$ is constant and $a_j$ parameters change} Let $N_{a}$ denote the number of different possible values of the cumulated Euclidean distance between any two  DC sequences of length up to 3, when varying $a_j$ values. 
\begin{itemize}
	\item Step 1: When emanating from the same state, the transitions are labeled by all the possible systematic symbols. This corresponds to ${\binom{q}{2}}$ different possible sub-terms in the Euclidean distance. When varying the $a_j$ parameters, if all combinations of $q^2$ parities labeling any couple of transitions are spanned, then the ensemble of generated codes covers all possible parity symbol combinations. This can be verified by taking one reference transition  (systematic and parity values kept constant) and by validating that the second transition can span all possible parity values, even when the systematic part is constant. In Fig. \ref{Fig:DC3Stages}, for the same starting state $E_i$, if transition ($s_i$, $p_i$) is taken as a reference, then parity $p'_i$ should be able to span the $q-1$ possible values (excluding $p_i=p'_i$) when $s'_i$ is kept unchanged. From (\ref{EqnRelStates}) and (\ref{EqnParityP2}), we have:
	  \begin{equation}
	  p_i+p'_i=a_2 (s_i+s'_i)
	  \label{Eq:ParAddition}
	  \end{equation}
	When $a_2$ spans  GF$(q)$, $p_i+p'_i$ takes the $q-1$ different non zero values. Therefore, $p'_i$ can actually take any possible value in $\{$GF$(q)\backslash\{p_i\} \}$. Consequently, the contribution of step 1 to $N_{a}$ is ${\binom{q}{2}}^2$.
	\item Step 2: With code structure $S_2$, $q^2$ different competing transitions are considered in this step regardless of the code parameters $a_j$, since any two transitions differ at least by their systematic or parity value. Therefore, the contribution of step 2 to $N_{a}$ is $\binom{q^2}{2}$. 
	\item Step 3: Again, dictated by the code structure, by exploiting the existing symmetry with regards to step 1, the contribution of step 3 to $N_{a}$ is ${\binom{q}{2}}^2$.
\end{itemize}	
In summary, $N_{a}={\binom{q^2}{2}}\cdot{\binom{q}{2}}^4$. 

The comparison of  $N_{\mu}$ with $N_{a}$ allows us to determine if the mapping plays a role in  the definition of the best code distance spectrum.  ${{\binom{q}{2}}}^4$ is a common factor of $N_{\mu}$ and $N_{a}$, it is then sufficient to compare the terms ${{\binom{q^2}{2}}}$ and ${\left ({\binom{q}{2}}+1\right )^2-1}$:

\begin{equation}
\begin{gathered}
\delta_N={ {\binom{q^2}{2}}}-\left ({\binom{q}{2}}+1\right )^2+1\\
= \dfrac{q^2 (q^2-1)}{2}-\dfrac{q(q-1)(q^2-q+4)}{4}\\
=\dfrac{q(q-1)^2 (q+4)}{4} > 0 \textrm{  }\forall \textrm{  } q>0
\end{gathered}
\label{CompN}
\end{equation}

Since $\delta_N > 0$, then $N_{a} > N_{\mu}$. In conclusion, varying the code parameters $a_j$ is sufficient to find the best code using the proposed search procedure. Therefore, in the rest of the paper, the mapping function $\mu$ is kept constant and is given in Table \ref{tab:QAM_mapping} for 16-QAM and 64-QAM constellations. It was chosen such that the binary image of the constellation symbol follows a Gray mapping. The binary images of the symbols are denoted by $b_3b_2b_1b_0$ and $b_5b_4b_3b_2b_1b_0$ for 16-QAM and 64-QAM, respectively, with the highest index corresponding to the most significant bit of the symbol representation.

Using the NB convolutional code structure $S_2$ depicted in Fig. \ref{Fig:GFConvCode}, the first terms of the distance spectra for all possibles codes in GF(16) and GF(64) were determined by varying coefficients $a_1$, $a_2$ and $a_3$, according to the methodology proposed  in Section \ref{sec:SearchMethod}. 
The primitive polynomials used to generate the elements of the Galois fields are: $P_{\text{GF}(16)}(D)=1+D^3+D^4$ and  $P_{\text{GF}(64)}(D)=1+D^2+D^3+D^5+D^6$.

	\begin{table}[htbp]
		\begin{center}
			\caption{Binary mapping of the in-phase $I$ and quadrature $Q$ axes for 16- and 64-QAM. }
			\begin{tabular}{||c|c|||c|c||} 
				\hline
				\multicolumn{4}{||c||}{16-QAM}\\
				\hline
				$Q$ value & $b_3b_1$ & $I$ value & $b_2b_0$  \\
				\hline\hline
				+3 & 00 & +3 & 00 \\
				\hline
				+1 & 01 & +1 & 01 \\
				\hline
				-1 & 11 & -1 & 11 \\
				\hline
				-3 & 10 & -3 & 10 \\
				\hline\hline
				\multicolumn{4}{||c||}{64-QAM}\\
				\hline
				$Q$ value & $b_5b_3b_1$ & $I$ value & $b_4b_2b_0$  \\
				\hline\hline
				+ 7 & 000 &  + 7 & 000 \\
				\hline
				+5 & 001 & +5 & 001 \\
				\hline
				+3 & 011 &  +3 & 011\\
				\hline
				+1 & 010 &  +1 & 010\\
				\hline
				-1 & 110 & -1 & 110 \\
				\hline
				-3 & 111 & -3 & 111 \\
				\hline
				-5 & 101 & -5 & 101 \\
				\hline
				-7 & 100 &  -7 & 100\\
				\hline
			\end{tabular} 			
			\label{tab:QAM_mapping}
		\end{center}
	\end{table}
	
Table \ref{tab:GFTable} provides the values of coefficients $a_1$, $a_2$ and $a_3$ for three specific codes resulting from the search in  GF$(16)$ and in GF$(64)$. In each field, C$_1$ is a code instance showing the worst distance spectrum, C$_3$ is a code instance showing the best distance spectrum, and code C$_2$ is a code with a ``medium'' distance spectrum. Table \ref{tab:GFTable} also displays the corresponding distance spectra truncated to the first two minimum distances  $d_1$ and $d_2$, with their multiplicities  $n(d_1)$ and $n(d_2)$, i.e. the number of DC sequences with distances $d_1$ and $d_2$.

		\begin{table}[htbp]
		\begin{center}
			\caption{Three representative codes obtained from the search over GF$(16)$ and GF$(64)$, with the two first terms of the squared Euclidean distance spectra $d_1^2$ and $d_2^2$ and the corresponding multiplicities  $n(d_1)$ and $n(d_2)$.}
			\begin{tabular}{|| c || c | c | c ||} 
				\hline
				\multicolumn{4}{||c||}{GF(16)}\\
				\hline
				Code & C$_1$ & C$_2$ & C$_3$\\
				\hline
				$(a_1,a_2,a_3)$ & $(12,4,0)$ & $(10,12,3)$ & $(13,7,11)$\\
				\hline\hline
				$d_1^2$ ($d_{min}^2$) & 1.20 & 2.00 & 4.00\\
				$n(d_1)$ & 22128 & 5532 & 22484\\
				\hline
				$d_2^2$  & 1.60 & 2.40 & 4.80\\
				$n(d_2)$ & 16596 & 8424 & 141144\\
				\hline\hline
				\multicolumn{4}{||c||}{GF(64)}\\
				\hline
				Code & C$_1$ & C$_2$ & C$_3$\\
				\hline
				$(a_1,a_2,a_3)$ & $(41,2,0)$ & $(41,1,24)$ & $(31,5,18)$\\
				\hline\hline
				$d_1^2$ ($d_{min}^2$) & 0.38 & 1.14 & 1.52\\
				$n(d_1)$ & 238422 & 1542390 & 652698\\
				\hline
				$d_2^2$  & 0.57 & 1.23 & 1.61\\
				$n(d_2)$ & 230886 & 4111444 & 1084014\\
				\hline
			\end{tabular} 			
			\label{tab:GFTable}
		\end{center}
	\end{table}
	
\section{Max-Log-MAP decoding of NB convolutional codes}\label{sec:MinLogMAPNB}
	In this study, an extension of the Max-Log-MAP decoding algorithm \cite{ScaledMLM00} to the symbol domain is used. The same decoding algorithm was applied in \cite{LivaNBTC1} and \cite{LivaNBTC2}.
	
	Similarly to the binary case, a forward recursion process is introduced to calculate the forward state metric $\alpha_i (j)$ corresponding to a state $E_i=j$ at trellis stage $i$:  	
	\begin{equation}\label{Eq:Alpha}
	\alpha_i (j)= \underset{j'\in{\{0 \cdots q-1\}}}{\max} (\alpha_{i-1} (j')+\gamma_{s, i-1} (j',j) +\gamma_{p, i-1} (j',j)) 
	\end{equation}
	where $\alpha_{i-1} (j')$ corresponds to the forward state metric at trellis stage $i-1$ for state $E_{i-1}=j'$. The systematic and parity transition metrics between states $E_{i-1}=j'$ and $E_{i}=j$ are represented by $\gamma_{s, i-1} (j',j)$ and  $\gamma_{p, i-1} (j',j)$ respectively, where $s(j',j)$ and $p(j',j)$ represent the corresponding systematic and parity values. 
	Similarly, the backward state metric related to state $E_i=j$ at trellis stage $i$ is computed through the backward recursion as: 
	\begin{equation}\label{Eq:Beta}
	\beta_i (j)=\underset{j'\in{\{0 \cdots q-1\}}}{\max} (\beta_{i+1} (j')+\gamma_{s, i} (j,j') +\gamma_{p, i} (j,j')) 
	\end{equation}		
	Then, for each symbol $u_k\in$ GF$(q)$, the logarithm of the probability that symbol $s_i$ is equal to $u_k$ at trellis stage $i$ denoted by $L_i(u_k)$ is computed by:
	\begin{equation}\label{Eq:HardDec}
	\begin{gathered}
	L_i(u_k)= \underset{(j,j') \in{\{0 \cdots q-1\}^2}\, | \, s(j,j')=u_k}{\max} \left[\alpha_{i} (j) +\beta_{i+1} (j')\right.\\\left.+\gamma_{s, i}(j,j')+\gamma_{p, i} (j,j')\right]
	\end{gathered}
	\end{equation}
	Hard decision is applied to identify the decoded frame, where $s_i=u_v \mid L_i(u_v)=\max (L_i(u_k))$.

	\section{Simulation Results} \label{sec:Results}

\begin{figure}[htbp]
		\centering
		\includegraphics[width=0.48\textwidth]{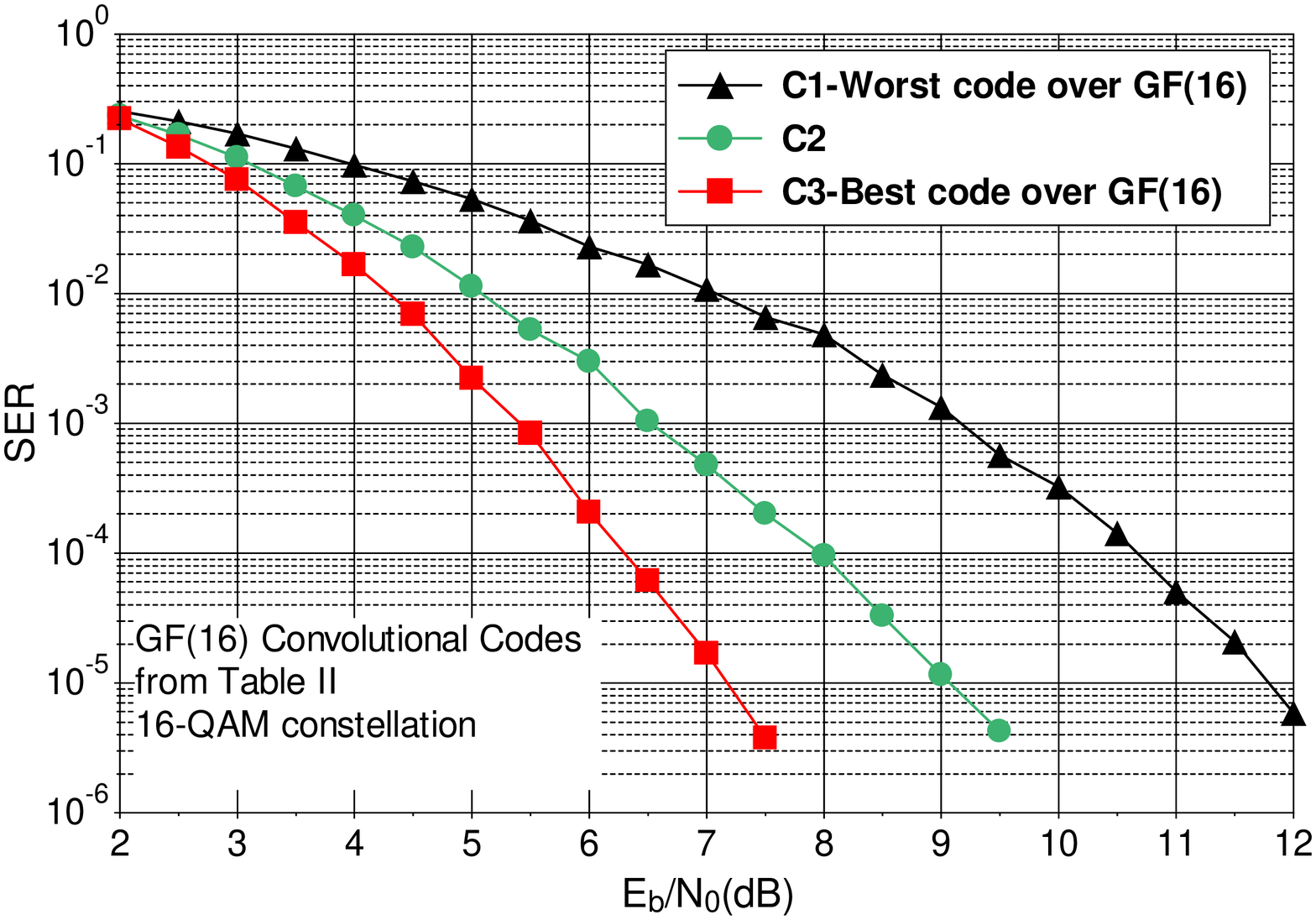}
		\caption{Performance comparison of convolutional codes over GF$(16)$ in terms of symbol error rates over the AWGN channel.}
		\label{GF16}
	\end{figure}

\begin{figure}[htbp]
		\centering
		\includegraphics[width=0.48\textwidth]{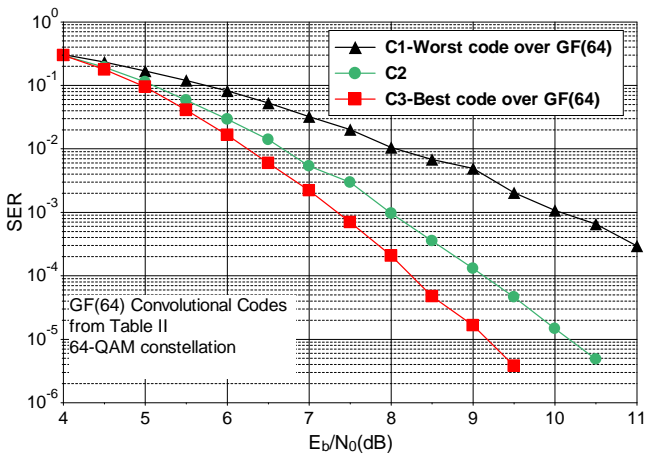}
		\caption{Performance comparison of convolutional codes over GF$(64)$ in terms of symbol error rates over the AWGN channel.}
		\label{GF64}
	\end{figure}

The error rate performance of the selected codes was assessed through Monte Carlo simulations over a Gaussian channel. We simulated the transmission of blocks of 100 symbols, corresponding to 400 bits in the case of GF(16) and 600 bits in the case of GF(64). The symbol error rate curves are shown in  Figure~\ref{GF16}  for the codes in GF$(16)$ and in Figure~\ref{GF64}  for the codes in GF$(64)$. These figures show the range of performance  that can be obtained with the proposed code structure and search process. We have also checked that several code instances with the same distance spectra display the same error rate performance (which is not shown in this paper for the sake of concision).  
		
\begin{figure}[htbp]
	\centering
	\includegraphics[width=0.48\textwidth]{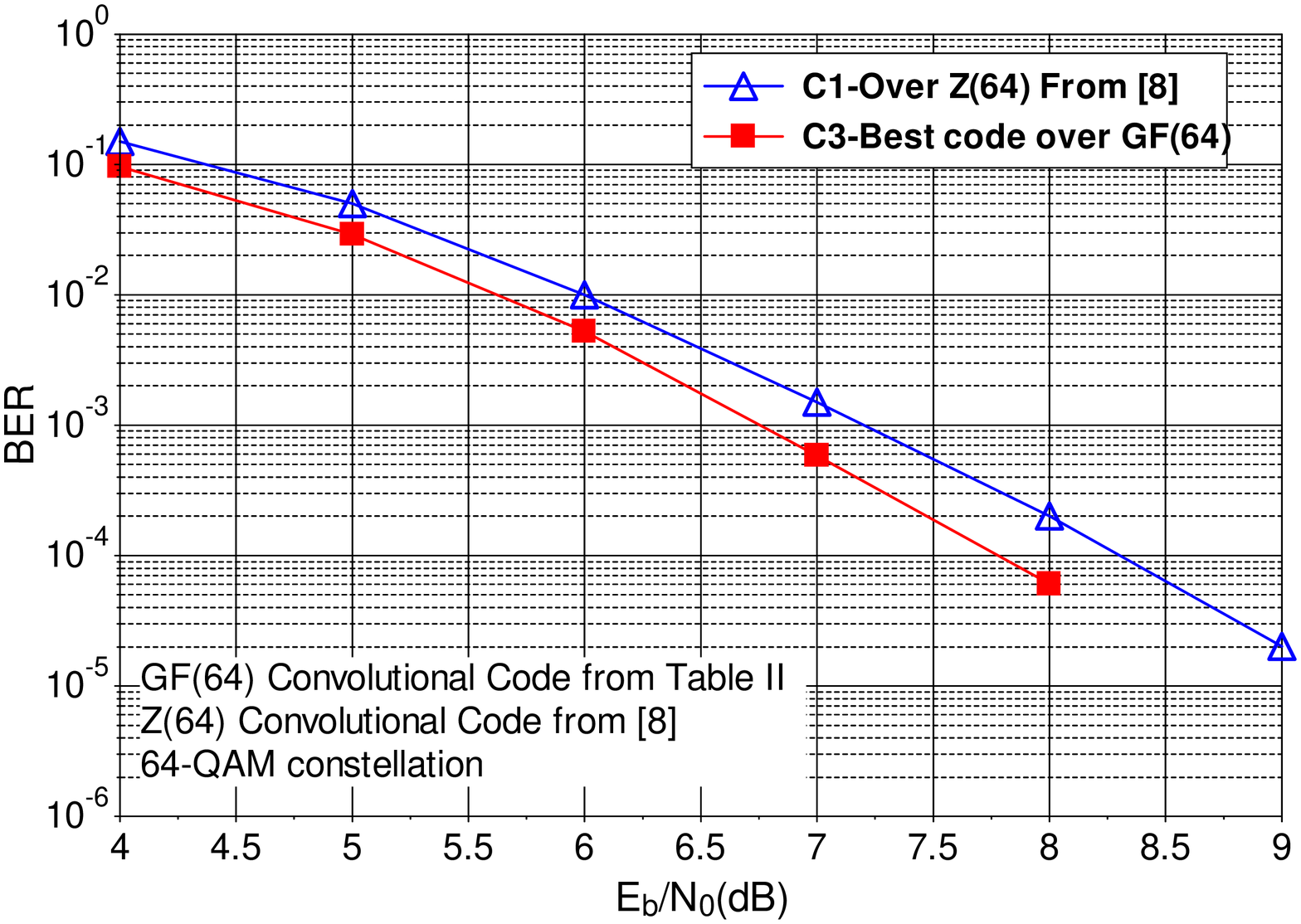}
	\caption{Performance comparison of the proposed convolutional code defined over GF$(64)$ (code C$_3$) with the code defined over $\mathbb{Z}_{64}$ and labeled C$_1$ in \cite{konishi2014coded}, in terms of bit error rate. Transmission is over an AWGN channel using 64-QAM constellation.}
	\label{GF64CompareWithKonishi}
\end{figure}

In \cite{konishi2014coded}, the construction of non-binary convolutional codes over rings was proposed and assessed in the case of AWGN channel. For the code defined over $\mathbb{Z}_{64}$, the resulting coded symbols are directly mapped to a 64-QAM constellation. Fig. \ref{GF64CompareWithKonishi} shows the comparison in terms of bit error rate of our best code over GF$(64)$ (C$_3$), with the best code defined in \cite{konishi2014coded} over $\mathbb{Z}_{64}$. This comparison shows that our best code in GF$(64)$  outperforms by around 0.5 dB the best proposed code over $\mathbb{Z}_{64}$.

	\begin{figure}[htbp]
		\centering
		\includegraphics[width=0.48\textwidth]{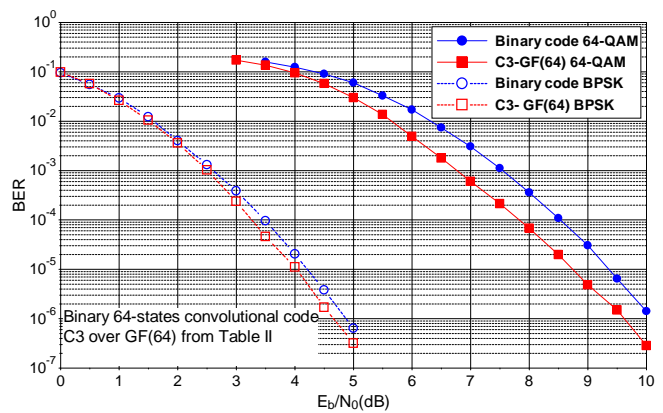}
		\caption{Performance comparison of the proposed convolutional codes over GF$(64)$ with the best known 64-state-binary code in terms of bit error rate. Transmission is over an AWGN channel, using BPSK and 64-QAM constellations.}
		\label{BinComp}
	\end{figure}
In addition, the designed NB code over GF(64) was compared with the binary 64-state recursive systematic convolutional code with generator polynomials $\left( 1,\frac{171}{133}\right)$ in octal. The latter is known to be the 64-state binary recursive systematic convolutional code with the highest minimum Hamming distance \cite{frenger1999convolutional}. Bit error rate curves were plotted  for both BPSK and 64-QAM constellations. 	
The curves in \mbox{Fig. \ref{BinComp}} show that, when combined with a 64-QAM, the designed NB code yields a gain in the order of 0.7 dB in comparison with the binary code, which is in accordance with the gain predicted by the capacity comparison in Fig.~\ref{CapacityFig} at coding rate $1/2$. In addition, when used with a BPSK modulation, the NB code still shows a slightly better performance than the binary code, although no capacity gain was expected.

	\section{Conclusion} \label{sec:Disc}

This paper presents a general framework for the design of recursive systematic convolutional codes defined over high-order Galois fields. A new low-complexity structure of convolutional codes is proposed, using only one memory element. This  structure allows the search space for the code to span all the combinations of systematic and parity symbols for each transition between two states in the trellis. It was also shown that the distance properties of the resulting codes are independent of the constellation mapping. The designed codes offer better performance than the non-binary codes previously proposed in the literature. They also outperform their binary counterparts when combined with their corresponding QAM modulation or with lower order modulations.
This study can be considered as a first step to design non-binary turbo codes. The exploration of interleaving techniques, puncturing patterns  and the reduction of the decoding complexity can be identified as the next steps to complete this work.

\section*{Acknowledgment} 
This work was partially funded by the EPIC project of the EU's Horizon 2020 research and innovation programme under grant agreement No. 760150, by Orange Labs and by the Pracom cluster. It has also received support from the PHC CEDRE program.

	\bibliographystyle{ieeetran}
	\bibliography{Biblio}
\end{document}